\documentstyle[10pt,epsfig]{article}

\begin{document}
\begin{center}

{\bf {Shifting of  the magnetic resonance peak to lower energy 
 in the superconducting state of underdoped YBa$_2$Cu$_3$O$_{6.8}$ } }
\vspace{0.2 in}

 P. Bourges$^1$, L.P. Regnault$^2$, Y. Sidis$^1$, 
J. Bossy$^3$ P. Burlet$^2$,\\ C. Vettier$^4$, J.Y. 
Henry$^2$ and M. Couach$^2$\\

\end{center}

\vspace{0.1in}

\begin{tabular}{lp{12.5 cm}}
(1) & {Laboratoire L\'eon Brillouin, CEA-CNRS, CE Saclay, 91191 Gif sur Yvette,
France}\\
(2) & {CEA Grenoble, D\'epartement de Recherche Fondamentale sur 
la mati\`ere Condens\'ee, 38054 Grenoble cedex 9, France}\\
(3) & {Institut Laue-Langevin, 156X, 
38042 Grenoble Cedex 9, France}\\
(4) & {European Synchrotron Research Facility, BP 220, 38043 Grenoble cedex,
France}\\
\end{tabular}

\vspace{.5in}

\noindent
{\bf Abstract.} - Inelastic neutron scattering has been used to determine 
the dynamic spin 
fluctuations in an underdoped  high temperature superconductor 
YBCO$_{6.8}$ single crystal. The magnetic resonance, 
that occurs around 40 meV in overdoped samples, is shifted to a lower 
energy, $E_r$= 34 meV. A constant ratio, $E_r/ k_B T_C= 4.9 \pm 0.2$, 
almost independent of the doping level,  is found.  According 
to numerous theoretical approaches, this finding  supports the idea 
that the resonance energy is proportional (approximately twice) to 
the superconducting gap. 

\vskip 0.5 cm

\noindent 
74.20.Mn Nonconventional mechanisms (spin fluctuations, polarons, and
bipolarons, resonating valence bond model, anyon mechanism, 
 marginal Fermi liquid, Luttinger liquid, etc.)\\
25.40.Fq Inelastic neutron scattering\\
74.72.Bk Y-based cuprates\\

The electronic correlations in high-$T_C$ superconductors are crucial 
in understanding normal as well as superconducting (SC) properties. 
They have been experimentally  evidenced by the persistence of spin 
fluctuations in the metallic state of cuprates. Especially, inelastic
 neutron scattering (INS) experiments have allowed to investigate 
the wavevector, energy and the doping dependencies of the imaginary 
part of the generalized  magnetic susceptibility ($Im \chi$) on a 
unique sample of YBa$_2$Cu$_3$O$_{6+x}$ (YBCO) in its metallic 
state\cite{rossat,sandiego,lpr}. First, $Im \chi$  is still located 
at the antiferromagnetic (AF) wavevector, 
$Q_{AF}=(1/2,1/2,q_l) \equiv (\pi,\pi)$. 
Second, in the SC state, when going from the weakly-doped regime 
$x \simeq 0.5$ to fully oxidized sample $x \simeq 1$\cite{lpr}, $Im \chi$ 
is gradually restricted to a smaller energy range. It is shifted to higher 
energies with increasing $x$ by a doping dependant  energy 
gap\cite{rossat,sandiego,lpr,bourges}, called {\it spin-gap}. 
In optimally doped samples and overdoped samples ($x \ge 0.94$),
it is now  well established that a strong sharp magnetic peak appears 
around 40 meV, called {\it resonance} by J. Rossat-Mignod {\it et 
al}\cite{rossat} for its first evidence.  Its magnetic origin has been 
then confirmed by  using  polarized neutron scattering\cite{mook}. 
Its major characteristic is that it disappears above 
$T_C$\cite{rossat,lpr,bourges,fong}. 
This peak also occurs  in  $Im \chi$ in addition to another contribution
which stands in the normal state\cite{rossat,lpr}. 
However, this  part of  $Im \chi$ is so reduced with increased 
doping\cite{lpr,bourges} that it becomes hardly
detectable\cite{bourges,fong,jltp} in fully oxydized samples.  

The  origin  of this magnetic resonance has been extensively discussed, 
especially in itinerant magnetism 
descriptions\cite{reso_wces,scgap+vhs,demler,shiba,reso_flora,levin,yakovenko}.
Beyond all the different hypotheses in these approaches, one can  basically  
consider that an enhancement of the electronic spin susceptibility arises 
at a specific energy due to the opening of the SC gap. This is usually  
argued to support a d$_{x^2-y^2}$ symmetry for the SC order parameter, $\Delta_k$\cite{reso_wces,scgap+vhs,demler,shiba,reso_flora,levin,flora}, 
ruling out other symmetries\cite{shiba} like extended s-wave\cite{reso_flora}. 
For others\cite{yakovenko}, the neutron data indicate that $\Delta_k$ has a 
{\it s}-wave symmetry and opposite signs in bounding and  antibounding 
electron bands formed within the (CuO$_2$)$_2$ bilayer.  More particularly,
a resonance peak  may occur  i)  due to a dynamic nesting\cite{reso_wces}, 
 ii) as a consequence of van Hove singularity  
in the electronic density of states (DOS)\cite{scgap+vhs}, iii)  because of 
a collective mode in the particle-particle channel  in the Hubbard 
model\cite{demler}. Other approaches propose that this resonance results
from spin-flip charge carriers excitations across the SC 
gap\cite{shiba,reso_flora,levin,yakovenko,flora}. In this view, 
the resonance energy, $E_r$, closely reflects the amplitude of the SC 
gap as well 
as its doping dependence. However, the behaviour of the resonance peak in the 
underdoped regime remains unclear. So far, only short reports of a 
previous investigation at $x\simeq 0.83$\cite{sandiego,lpr} 
have emphasized that crucial point.

In this letter, we establish that, for an underdoped sample exhibiting a 
$T_C$= 82 K, the magnetic resonance peak is shifted to $E_r$= 34 meV. This
energy leads to a resonance energy and $T_C$ ratio  
which is almost the same as in optimally doped and overdoped samples,
$E_r/ k_B T_C= 4.9 \pm 0.2$. It  confirms the following picture in which
the resonance energy seen in  INS experiments actually  
measures the amplitude of the SC gap. 

The sample used in which the oxygen content can be changed was the same as in
our previous investigations\cite{rossat,sandiego,lpr}. It has been prepared by 
heat treatment to achieve a nominal oxygen content x = 0.8. Macroscopic 
susceptibility measurements have been performed to determine  $T_C$:
when applying a small magnetic field (H=0.1 Oe), an onset superconducting 
transition temperature occurs at 87 K (midpoint at 82 K).
A small amount of the 60 K-$T_C$ phase ($\le$ 10 \%) can be estimated
from these measurements.

Unpolarized inelastic neutron scattering experiments have been performed  
on the triple axis IN8 at the Institut Laue Langevin (ILL-Grenoble). 
Cu111 reflection was chosen as monochromator and horizontally focussed 
pyrolytic graphite PG002 as analyzer. We worked without collimators except 
between the analyser and the detector,  where a 60' Soller slit has been 
put to reduce the background level. A pyrolytic graphite filter was placed 
in the scattered beam to avoid higher order contaminations in $k_F$-fixed 
mode.  The  sample was mounted inside  an orange-type cryostat in the 
scattering plane usual in such investigations\cite{bourges} where the
 (110) and (001) directions are lying  within the horizontal plane.
We have worked with a final wavevector $k_F= 4.1$ \AA$^{-1}$, yielding 
a full width at half maximum (FWHM) of the resolution  along
the  (110) direction (see Fig. 2.b.)
of $\Delta ^{resol}q \simeq$  0.1 r.l.u. for energy transfers of 
 interest,  $\hbar\omega \simeq$ 35 meV.  It yields an energy resolution 
width of  4.7 meV also for $\hbar\omega \simeq$ 35 meV.
 As we have to measure a broad 
 in q-space but sharp in energy  signal,  such a spectrometer resolution is 
 particularly well adapted.

The major experimental problem encountered in such INS experiments 
is the extraction of the magnetic contribution from the scattering 
arising from the nuclear lattice:  
There is no unambiguous determination of the  energy dependence of 
$Im \chi$ over a wide range of energy. For instance, 
the use of polarized neutron\cite{mook} has even led to spurious effects
like the persistence of the resonance peak in the normal state.
All unpolarized\cite{rossat,sandiego,lpr,bourges,fong,jltp} neutron 
measurements have shown its disappearance above $T_C$. 
However, few complementary methods as well as  the accumulation of 
neutron data allow to solve the experimental 
problem\cite{lpr,bourges,mook,fong,jltp}. The major criteria 
to determine the magnetic signal are  the dependencies of the
measured intensities on the wavevector as well as the
temperature\cite{lpr,bourges,jltp}. Therefore,  we performed 
q-scans along the (110) direction with as many fixed energy transfers 
as possible. The $q_l$ value along the (001) direction 
was chosen to get the maximum of the magnetic structure factor
 ($q_l$-dependence is  reported in Ref. \cite{jltp}). 
For instance, Fig. 1.a displays the neutron intensity obtained 
at $\hbar\omega$= 35 meV and  $q_l$= 5.2. It  shows a well-defined 
maximum at q=0.5, corresponding to the AF wavevector ($\pi,\pi$), 
at all temperatures investigated from 4 K up to 200 K. However, 
the correlated signal significantly decreases with heating. 
Q-scans having similar {\bf q}- and  T-dependencies have been 
performed at few other energy transfers. At 39 meV (for details 
see \cite{jltp}), correlated intensity is reduced on increasing the 
scattering  wavevector  {\bf  Q}, according to the expected magnetic 
form factor of copper spins.  Magnetism is indeed 
known to decrease on increasing {\bf  Q} in contrast to the 
scattering arising from phonons\cite{lovesey}.
Such a wavevector and  temperature dependencies demonstrate the 
magnetic origin of this signal.

The next step is to determine the energy profile of the spin susceptibility.
Schematically, the energy scan performed at the AF wavevector is the 
sum of nuclear and magnetic contributions, and then  consists
of three terms as:

\begin{equation}
I(Q_{AF},\omega)= A + I_{nuclear}(\omega) + I_{magnetic}(\omega) 
\end{equation}

where A is a constant independent of T and $\omega$. Both, nuclear 
and magnetic contributions follow temperature population factor 
according to the  detailed balance\cite{lovesey}:

\begin{eqnarray}
I_{nuclear}(\omega) =  B(\omega) \{ 1+n(\omega, T)\} \nonumber \\
I_{magnetic}(\omega) = Im\chi(Q_{AF},\omega) \{ 1+n(\omega, T)\}
\end{eqnarray}

Because the spin fluctuations contribution  at 200 K is weak 
(see for instance Fig. 1.a), the energy scan performed at the 
AF wavevector, $Q_{AF}$=(0.5,0.5,5.2), mainly reflects,
at that temperature, the shape and the amplitude of the nuclear 
contributions, $B(\omega)$. Then, the q-scans performed at few 
constant energies crosscheck that determination and allow to take 
into account weak 200 K magnetic contribution in a self-consistent 
way. Once this nuclear part, $B(\omega)$, has been determined, 
the magnetic part, $Im\chi(Q_{AF},\omega)$, is then simply obtained
by difference (Eqn. (1)) and temperature correction (Eqn. (2)).

As shown in Fig 2.a, the spin susceptibility is spread over a wide 
energy range. This is in contrast with the overdoped  regime, in which 
$Im \chi(Q_{AF},\omega)$  is restricted to a narrow energy  range 
around 40 meV\cite{lpr,bourges,fong}. However, $Im \chi$ is 
characterized by a clear maximum at 34 meV which might be a 
resonance peak. We have fitted the q-scans performed at different energy 
transfer by a simple Gaussian lineshape. The deduced q-width, 
$\Delta q$, is reported in Fig 2.b versus energy. Magnetic correlations 
amplitude is maximum (Fig. 2.a) when they are sharper in q-space 
(Fig. 2.b). $\Delta q$ displays a minimum around 35 meV, $\Delta q$= 
0.14 r.l.u., which is still significantly broader than the
resolution width. In YBCO$_{6.97}$\cite{bourges}, similar energy 
dependence and values of $\Delta q$  have been found between 35 meV 
and 45 meV. In terms of correlation lengths, it yields $\xi_{E_r}$ = 
10 \AA\ after a proper deconvolution from the spectrometer resolution. 
Actually, such a characteristic length associated with the resonance 
is found to be independent from the doping\cite{lpr}. 
Fig. 3 compares the temperature 
dependence of the neutron intensity  obtained at 35 meV and 
at 39 meV. A change in the slope is observed at $T_C$ at the former 
energy whereas only a smooth behaviour is found at the latter energy.
Such a temperature dependence of the 34 meV peak also indicates
its resonating nature. Therefore, disappearing above $T_C$ (Fig. 2.a), 
the 34 meV peak in underdoped YBCO$_{6.8}$ has the same physical 
origin as the magnetic resonance peak around 40 meV in optimally and  
overdoped samples, i.e. for x $\ge$ 0.92\cite{rossat,lpr,bourges,mook,fong}. 

Apart from the resonance peak, $Im \chi(Q_{AF},\omega)$ displays other
important features. It is limited at low energy by a spin gap
defined by the inflexion point at $E_G = $ 20 meV. Between this 
energy gap and the resonance peak, there is a plateau at 25 meV
(Fig. 2.a). This part of $Im \chi$ is evidenced by the temperature 
dependence of the q-scan (Fig. 1.b). This plateau is characterized 
by a q-width different from the one of the 
resonance peak (Fig. 2.b). From the energy dependence of  $\Delta_q$ ,
the plateau extends to higher energies, i.e. above $E_r$. 
This correlated magnetic signal clearly remains up to the 
highest investigated energy, 55 meV.  In contrast, no magnetic 
fluctuations was sizeable at 47 meV in overdoped  
YBCO$_{6.97}$\cite{bourges}.  The amplitude 
of this non-resonant contribution  for energies 
around $E_r$ might be suggested  by continuity
(dashed lines in Fig 2.a and in Fig. 3). However,  only 
quantitative models allow to correctly distinguish  the
resonance intensity from this second contribution. 
After a Gaussian deconvolution of the q-width,  
this plateau  can be associated  with an AF correlation 
length, $\xi_{AF} = 5 \pm 1$ \AA\ , twice smaller than
 the one related to the resonance\cite{lpr}.
It is worth noting  that non-negligible  magnetic fluctuations 
corresponding to the non-resonant part 
clearly remain  in the normal state at all energies investigated (Fig. 2).
 Actually, the vanishing of the normal state 
spin fluctuations (which might be due to experimental limitations)
 only  occurs  in the overdoped regime\cite{bourges,fong}. 
This has led to controversies in the literature. 

Our results demonstrate that the resonance peak is shifted to lower 
energy in the underdoped regime. From the temperature dependence at 
39 meV (Fig. 3.b), its occurrence around 40 meV is ruled out.
Almost independently of the doping, the resonance energy and 
$T_C$ are barely proportional to each other\cite{keimer}
as $E_r/ k_B T_C= 4.9 \pm 0.2$. This mean ratio also stands  
 for overdoped samples in which both $T_C$ and $E_r$
are sligthly renormalized (from $E_r=$ 41 meV for optimally 
doped samples $x\simeq 0.93$ to 39 meV for YBCO$_7$)\cite{lpr}. According to 
numerous calculations\cite{shiba,reso_flora,levin,yakovenko,flora}, 
the resonance energy corresponds to roughly twice the 
SC energy gap (2 $\Delta_{max}$ for a d$_{x^2-y^2}$-wave symmetry),
i.e. $E_r$ would approximately be the energy necessary to 
break a Cooper pair. However, accurate calculations taking 
into account realistic electronic band show that this 
proportionnality factor between $E_r$ and $\Delta_{max}$ should be
renormalized to smaller values\cite{reso_flora,levin,flora,ops}, 
and that it might also be slightly doping dependent\cite{ops}.

Furthermore, the doping dependence of the characteristic magnetic resonance 
peak energy, $E_r$, enlights why the well-known anomalous 
behaviour  of  the 340 cm$^{-1}$ $B_{1g}$ phonon measured in 
Raman scattering\cite{thomsen} occurs only in fully 
oxydized samples\cite{altendorf}. Indeed, the phonon 
softening and broadening (abruptly occuring just below $T_C$) 
are currently interpreted in terms of a coupling of phonons with 
the electronic bath responsible of the superconductivity. 
Consequently, only phonons whose frequency is located close
to 2$\Delta_{max}$ (the SC gap has been estimated\cite{thomsen} 
to be  $2 \Delta$= 316 cm$^{-1}$ =39.2 meV for YBCO$_7$) 
should display marked modifications at $T_C$.
As discussed above, the resonance peak energy also probes the 
SC gap. Noticing the slight renormalization of its energy
when increasing the doping, this implies a shift of the SC gap,
and consequently, the vanishing of the phonon anomaly in non fully oxydized 
samples due to its extreme sensitivity with the 
frequency\cite{thomsen}. 

Finally, our results show that the neutron magnetic resonance peak is 
shifting to lower energies in YBCO$_{6.8}$ ($T_C$= 82 K).
However, a large part of $Im\chi$ remains in the normal state in 
this underdoped sample. The SC transition temperature does not affect 
this contribution. Then, the simple itinerant pictures described 
above cannot account for the complex behaviour of the spin 
susceptibility in cuprates. Other approaches involving the 
localized-itinerant duality of the magnetism, as  
in the framework of the  t-t'-J model\cite{flora}, have 
to be considered.

\section*{ACKNOWLEDGEMENTS}

We kindly acknowledge  the technical support of ILL, and especially 
D. Puschner. We also thank H.F. Fong, B. Hennion, B. Keimer, 
L. Manifacier, I. Mazin and F. Onufrieva for enlightening discussions
and/or critical reading of the manuscript.

\clearpage 

\begin{figure}

 \centerline{\epsfig{file=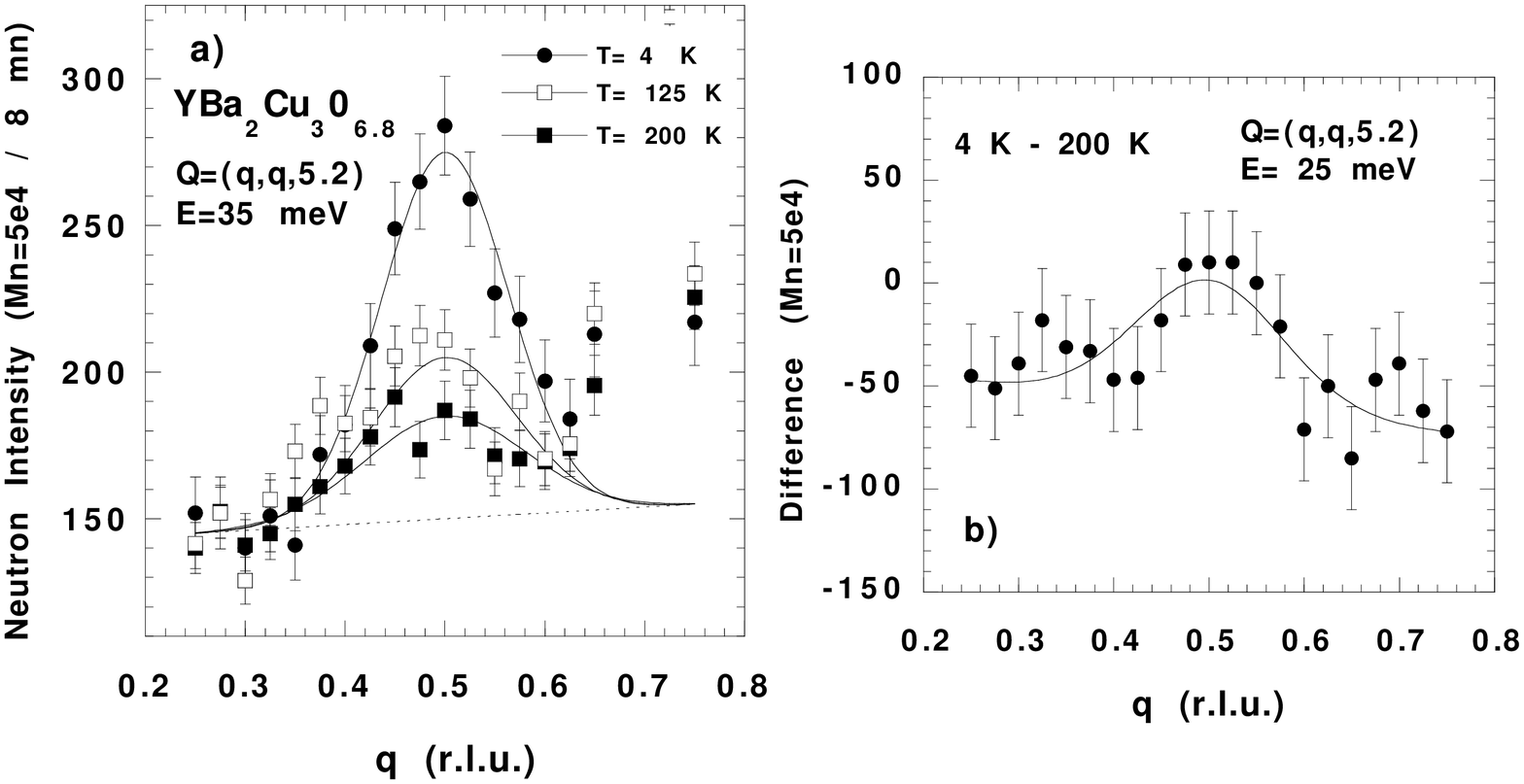,height= 8 cm,width= 15 cm}}
\caption{a) Q-scans across the magnetic line at an energy transfer of 35 meV at 
different temperatures. Data reported at 125 K have been 
measured independently at 100 K and 150 K, 
and then sumed up  to improve statistics. At 200 K, 20 counts have
been substracted at each point to scale the background to the scans measured
at lower temperature. The higher intensities located around q=0.7
correspond to a well-known accidental Bragg peak scattering[4].
b) Difference of q-scans measured at $\hbar\omega$= 25 meV
 between  4 K and 200 K. The full lines correspond to a fit to a 
Gaussian lineshape 
$I(Q,\omega)=Im \chi(Q_{AF}, \omega) \{ 1+n(\omega, T)\} \exp( -4 \ln 2
 \big( {{ q-q_{AF}}\over{\Delta_q}} \big)^2)$ with $q_{AF}= 0.5$
above a sloping background (dashed line).
}
 \label{fig:35_6.8} \end{figure}

\begin{figure}
 \centerline{\epsfig{file=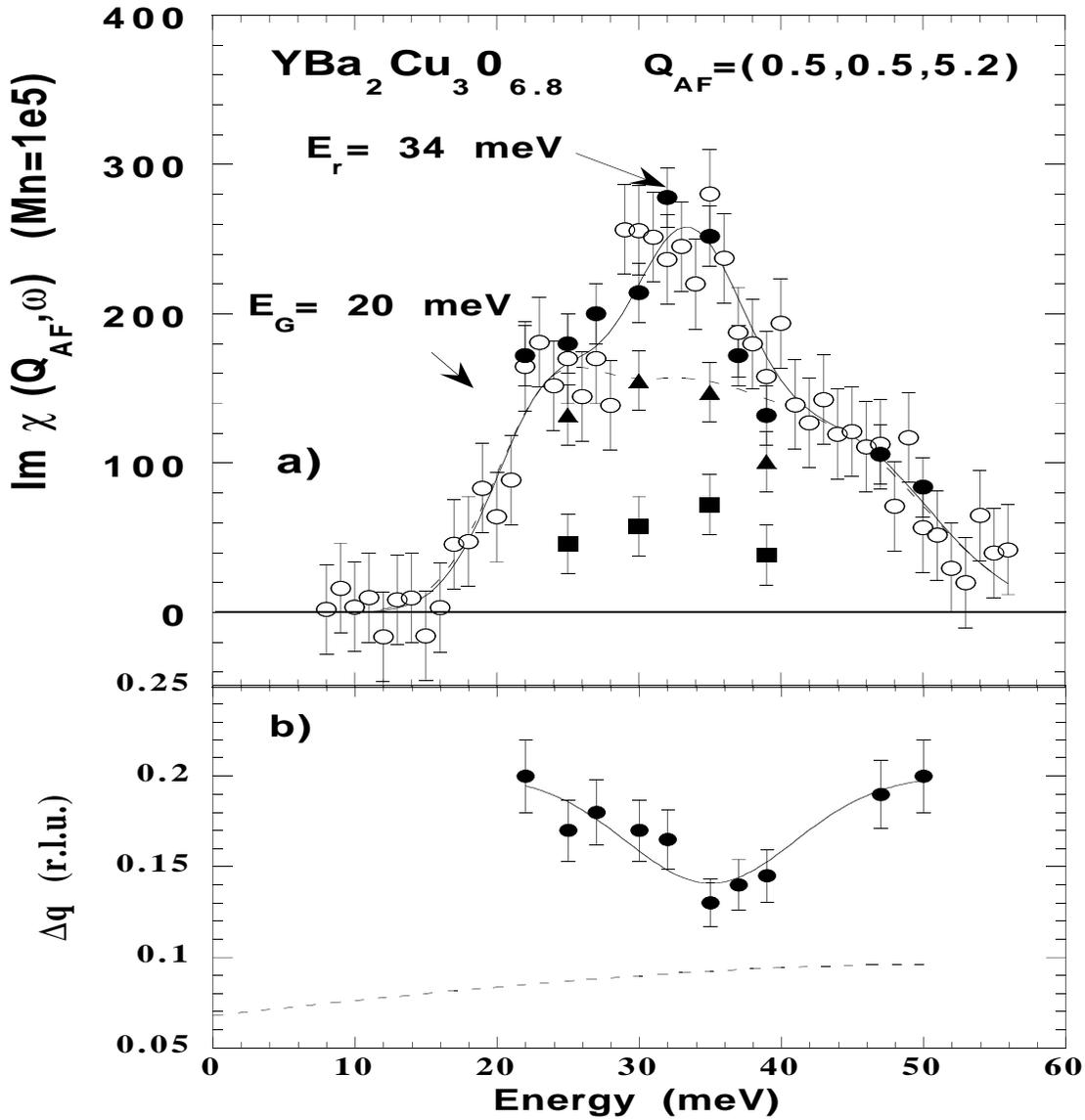,height= 15 cm,width= 15 cm}}

\caption{a) $\chi"(Q_{AF},\hbar\omega)$ at different temperature:
          circles at 4 K (SC state),  triangles at 100 K (normal state)
          and squares at 200 K. Closed symbols have been obtained 
          from q-scans (as in Fig. 1) at each energy.
         b) FWHM q-width,  $\Delta_q$ , in the (110) direction from 
          a  Gaussian fit of q-scans. The resolution q-width is represented  
          by the dashed line. Other lines are guides to the eye.}
\label{fig:w_6.8} \end{figure}

\begin{figure}

\centerline{\epsfig{file=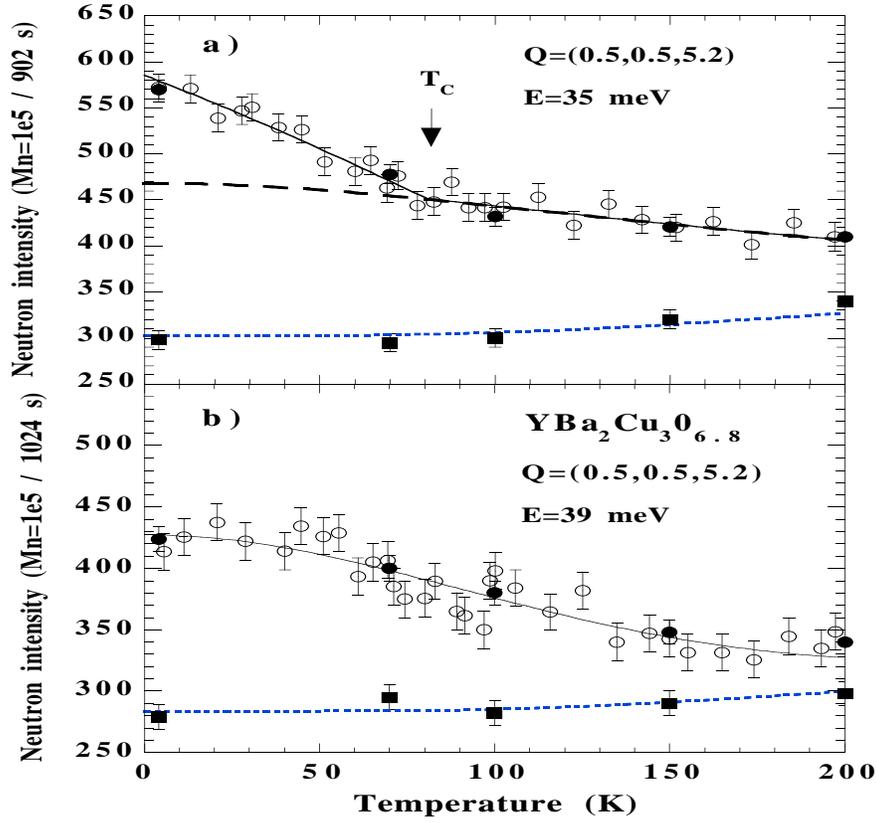,height=15 cm,width= 15 cm}}

\caption{Temperature dependence of the neutron intensity at two energies
 (circles). Closed circles have been obtained from q-scans (as in Fig. 1) at 
each energy whereas squares display the background deduced from them.
 Dotted lines correspond to a fit of the nuclear contributions by
 $I_{nuclear}=A + B(\omega) \{ 1+n(\omega, T)\}$. 
The full lines are guides to the eye. }
\label{fig:temp_6.8} \end{figure}

\end{document}